# Effects of brake wear nanoparticles on the protection and repair functions of the airway epithelium

Chloé Puisney-Dakhli [1], Evdokia K Oikonomou [2], Mickaël Tharaud[3], Yann Sivry[3], Jean-François Berret[2], Armelle Baeza Squiban[1]*

[1]: *Université Paris Cité, Unit of Functional and Adaptive Biology (BFA) UMR 8251 CNRS, F-75205, Paris, France.*
[2]: *Université Paris Cité, Laboratoire Matière et Systèmes Complexes, UMR 7057 CNRS, Paris, France*
[3]: *Université Paris Cité, Institut de Physique du Globe de Paris, CNRS, F-75005, Paris France*

**Abstract:** Long term exposure to particulate air pollution is known to increase respiratory morbidity and mortality. In urban areas with dense traffic most of these particles are generated by vehicles, via engine exhaust or wear processes. Non-exhaust particles come from wear processes such as those concerning brakes and their toxicity is little studied. To improve our understanding of the lung toxicity mechanisms of the nanometric fraction of brake wear nanoparticles (BWNPs), we studied whether these particles affect the barrier properties of the respiratory epithelium considering particle translocation, mucus production and repair efficiency. The Calu-3 cell line grown in two- compartment chambers was used to mimic the bronchial epithelial barrier. BWNPs detected by single-particle ICP-MS were shown to cross the epithelial tissue in small amounts without affecting the barrier integrity properties, because the permeability to Lucifer yellow was not increased and there was no cytotoxicity as assessed by the release of lactate-dehydrogenase. The interaction of BWNPs with the barrier did not induce a pro-inflammatory response, but increased the expression and production of MU5AC, a mucin, by a mechanism involving the epidermal growth factor receptor pathway. During a wound healing assay, BWNP-loaded cells exhibited the same ability to migrate, but those at the edge of the wound showed higher 5-ethynyl-2'-deoxyuridine incorporation, suggesting a higher proliferation rate. Altogether these results showed that BWNPs do not exert overt cytotoxicity and inflammation but can translocate through the epithelial barrier in small amounts and increase mucus production, a key feature of acute inflammatory and chronic obstructive pulmonary diseases. Their loading in epithelial cells may impair the repair process through increased proliferation.

# 1 - Introduction

It is now clearly recognized that air pollution is responsible for increased morbidity and mortality, whether it is after short- or long-term exposure (Landrigan et al., 2018). The composition of air pollution is complex, combining gases, vapors and particles, and it varies according to source, location and season (Kelly and Fussell, 2020). The particulate component is considered to have a major role in health effects. The International Agency for Research on Cancer sees outdoor air pollution as one of the major environmental causes of cancer deaths and has classified particulate matter (PM) as carcinogenic to humans (Group 1) (Loomis et al., 2013). Air quality standards are defined for $PM_{10}$ and $PM_{2.5}$, particles with an aerodynamic diameter of equal to or less than 10 or 2.5 µm, respectively, in order to limit their levels in outdoor air for environmental and health protection. According to World Health Organization guidelines, their annual mean concentrations should not exceed 15 µg/m$^3$ and 5 µg/m$^3$ for $PM_{10}$ and $PM_{2.5}$, respectively, but national air quality standards still exceed these values, emphasizing the need for further action on emission sources. Since vehicle emissions, such as from diesel engines, contribute to particle pollution in high-traffic areas, regulations imposing lower emissions have favored the development of cleaner technologies (Greim, 2019). By contrast, there were until recently no regulations regarding other types of emissions, such as those from braking and tire wear. However, the European Union has now released new Euro 7 standards to reduce their emissions (EU COM(2022) 586 final). It is therefore expected that the proportion of non-exhaust particles will increase due to the exhaust-related regulations as well as to the rise in the use of electric vehicles, which will also contribute to the production of braking and wear particles (Piscitello et al., 2021).

Exhaust particles are related to the combustion process in engines, meaning they are essentially made of elemental and organic carbon in association with inorganic substances (Salameh et al., 2015). Their adverse effects have been widely studied, underlining the role of organic compounds such as polyaromatic hydrocarbons (Cassee et al., 2013). Non-exhaust particles such as brake wear particles (BWPs), on the other hand, have a very distinct composition as they are rich in metals and have been the subject of only a limited number of toxicological studies to assess their effects (Gasser et al., 2009) (Gerlofs-Nijland et al., 2019) (Figliuzzi et al., 2020). BWPs were initially considered as being rather large in size and to contribute to the $PM_{10}$ fraction, but recent studies have highlighted the abundance of nanoparticles (Vojtíšek-Lom et al., 2021) (Gonet et al., 2021). In our previous work with BWPs recovered from vehicles in a test center, we identified a nano-size fraction that represented 26% by mass. Preliminary studies on a 2D human bronchial epithelial model showed the ability of brake wear nanoparticles (BWNPs) to induce a loss of viability associated with reactive oxygen species generation, but with limited pro-inflammatory effects (Puisney et al., 2018).

Inhaled particles are deposited all along the respiratory tract from the nose to the alveoli, according to their size (Oberdörster et al., 2005) In the tracheobronchial region, mucociliary epithelium constitutes a robust physicochemical barrier. The intercellular junctions prevent the paracellular passage of particles and particle clearance is affected by the joint action of mucus which traps particles, and ciliary beating, which moves mucus towards the oropharyngeal junction. Nevertheless, this epithelium is constantly exposed to airborne pollutants/pathogens that can compromise its barrier properties (Aghapour et al., 2022). It can repair itself thanks to the migration of undamaged cells and their proliferation and redifferentiation.

In order to improve our understanding of the toxicity mechanisms of the nanometric fraction of BWPs, we studied whether these particles could affect the barrier properties of the respiratory epithelium, considering particle translocation, mucus production and repair efficiency. For this purpose the Calu-3 cell line was grown in a 2-compartment chamber (Transwell™), allowing cell polarization, mucus production and formation of a tight epithelium (George et al., 2015b) (Sanchez-Guzman et al., 2021). This enabled us to investigate effects specific to the properties of the bronchial epithelial barrier that are not accessible when using a classic 2D culture, namely: i) NP translocation, ii) protection mechanisms via mucus production due to more advanced cell differentiation, iii) repair mechanisms after injury, which can be understood more realistically with a fully confluent and tight culture. The epithelial barrier was exposed for 24 hours to BWNPs at non-cytotoxic concentrations. Their effects were compared to those of $\gamma$-Fe$_2$O$_3$ nanoparticles ($\gamma$-Fe$_2$O$_3$ NPs) in order to mimic the iron fraction of BWNPs, as they were shown to be present in the nanosize particles released by automotive brakes (Kukutschová et al., 2011). To evaluate the translocation of BWNPs through the tight epithelium, single-particle inductively coupled plasma mass spectrometry (spICP-MS) was used to determine the number of iron-containing nanoparticles in the basolateral culture medium. Based on monitoring of the iron signal at relatively high frequency (*i.e.* dwell-time *ca.* a few milliseconds), this technique discriminates between dissolved iron and iron-bearing nanoparticles (Tharaud et al., 2017). Mucus protection as well as the ability of an epithelium previously exposed to BWNPs to repair through migration and proliferation were investigated studying MUC5AC expression and using the wound healing assay, respectively (Figure S1).

## 2 - Materials and methods

### 2.1-Nanoparticles

BWNPs were recovered from BWPs that were obtained from a motor vehicle testing center as previously described and characterized (figure S2) (Puisney et al., 2018). Briefly, after sonication, they were isolated by a combination of filtration and ultra-centrifugation techniques. Stable colloidal dispersions with a size distribution peaking at 66 nm and a zeta potential of -26 mV were obtained. The most abundant elements were copper and iron. For toxicological studies, the suspension was diluted in culture medium where NPs tend to agglomerate, exhibiting a most frequent mode at 320 nm (Puisney et al., 2018). Iron oxide nanoparticles were synthesized by alkaline co-precipitation of iron(II) and iron(III) salts (Massart et al., 1995). The salts were dissolved in a hydrochloric solution and co-precipitated with concentrated ammonia, leading to the spontaneous formation of Fe$_3$O$_4$ (magnetite) nanocrystals. Following several washes, the pH of the dispersion was decreased to 1.5 by addition of nitric acid. In a second stage, addition of a large excess of ferric nitrate at water boiling temperature led to the oxidation of magnetite into stable maghemite nanocrystals ($\gamma$-Fe$_2$O$_3$). The $\gamma$-Fe$_2$O$_3$ nanoparticles were sorted according to their size by successive steps of phase separation. For the batch used in the work, the dispersion stability was ensured by electrostatic repulsive interactions (Berret et al., 2006). It was characterized by vibrating sample magnetometry, transmission electron microscopy and dynamic light scattering (Berret, 2009). The $\gamma$-Fe$_2$O$_3$ nanoparticles considered had a median diameter of 13.5 nm and a dispersity of 0.26 (figure S3). The light scattering provided hydrodynamic diameters of 50 nm, probably indicating a slight aggregation.

### 2.2-Cell culture

Human lung adenocarcinoma Calu-3 cells (ATCC, HTB-55$^{TM}$, Manassas, VA, USA) were grown at 37 °C under 5 % CO$_2$ atmosphere in EMEM medium (Sigma Aldrich, Saint-Quentin Fallavier, France) supplemented with 10% fetal bovine serum (FBS, Eurobio, Les Ulis, France) and 1% non-essential amino acids (Sigma-Aldrich). Cells were trypsinized once a week (Trypsin-EDTA, Sigma-Aldrich). Passages 24 to 30 were used. After cellular expansion in T75-flasks (Costar, Corning, New York, USA), cells were grown in two compartment chambers. They were seeded at 500,000 cells/cm² in 500 µL of complete culture medium onto polycarbonate Transwell Filters (TF, Costar) with 3 µm pore size, and 1,500 µL of complete culture medium was added to the basolateral chamber. Media were changed every two days. A polarized and tight epithelial barrier was obtained after 14 days of culture (George et al., 2015).

### 2.3-Cell treatment

After 14 days of culture, TF-confluent Calu-3 cultures were grown without FBS for 24h. Cells were apically treated for 24 h with NPs at increasing concentrations from 1 to 100 µg cm$^{-2}$, corresponding to 3 to 300 µg mL$^{-1}$ for 24h. These concentrations are nominal concentrations. However, the deposited concentrations are probably lower as using a dosimetry model we previously observed that only a fraction of suspended BWNP are deposited on cells (Puisney et al., 2018). A cytotoxicity assay or an enzyme-linked immunosorbent assay (ELISA) was performed, on apical and basolateral media recovered after treatments and kept frozen until use. Cells were rinsed two times with PBS and fixed for immunolabelling and confocal microscopy observations. For the study of the EGFR pathway, inhibitors or antibodies were added to cell cultures 30 min before exposure to NPs. AG1478 (10 µM), an EGFR tyrosine kinase activity inhibitor (tyrphostin) and a non-specific tyrphostin (AG-9, 10 µM), were all from Calbiochem (Nottingham, UK). Anti-EGFR neutralizing monoclonal LA1 antibody used at 0.5 µg/mL was from R&D Systems (Abingdon, UK) and the monoclonal mouse non specific IgG antibody was from DakoCytomation, Trappes, France).

### 2.4-Lucifer yellow permeability

The Lucifer yellow (LY) paracellular permeability assay (Sigma-Aldrich) was performed according to the manufacturer's instructions. Briefly, LY (0.1 mg/mL in HBSSCa$^{2+}$/Mg$^{2+}$) was added to the apical (0.5 mL) compartment and the basal compartment was filled with 1 mL of HBSSCa$^{2+}$/Mg$^{2+}$. 100 µL of the apical and basal solutions was recovered after 45 min incubation for fluorescence measurement with excitation and emission wavelengths set at 485 nm and 538 nm. Calu-3 cells form a tight epithelium if permeability <2 %.

### 2.5-Cytotoxicity assay: LDH assay

After 14 days of culture, TF-confluent Calu-3 cultures were grown without FCS for 24 h before being exposed to NPs at increasing concentrations from 1 to 100 µg cm$^{-2}$, corresponding to weight concentrations 3 to 300 µg mL$^{-1}$ for 24 h. At the end of the exposure, apical and basal culture media were recovered and centrifuged at 3000 $g$, 4 °C for 10 min. LDH activity was measured in the supernatant using the CytoTox 96® Non-Radioactive Cytotoxicity Assay (Promega, Madison, USA) according to the manufacturer's protocol. The absorbance signal reflecting the LDH activity was measured at 490 nm with an ELx808 microplate reader (BioTek Instruments, Winooski, USA). The percentage cytotoxicity was estimated as (the absorbance of the sample/ the absorbance of the control)*100.

### 2.6-Wound healing assay, videomicroscopy and image analysis

After NP treatment, a linear scratch was made using a 100 µL tip and aspiration to suck up detached cells. By training, reproducible scratches of mean width l = 864 ± 38 µm were obtained. In order to track cells by videomicroscopy, they were labelled with CellTracker™ Green CMFDA (Thermo Fisher Scientific, Waltham, USA). Pictures were taken every 30 min over 72 hrs to follow the wound closure using a wide-field Leica Statif DMI 6000 equipped with an sCMOS Andor camera (plateforme ImagoSeine, Institut Jacques Monod) and with a CO$_2$ and temperature-controlled chamber. Image acquisition was controlled by MetaMorph software and ImageJ was used to process images.

### 2.7-Cell proliferation

To investigate proliferation during the wound healing assay, 5-ethynyl-2'-deoxyuridine (EdU) incorporation in cells in S phase was determined. 10 µM EDU in culture medium was added to cells just after the scratch for different times of incubation (1, 4, 6, 10, 16, 24, 48 and 72h). Cells were rinsed with PBS and fixed using 4% paraformaldehyde in PBS. After rinses, the « Click-iT™ EdU Alexa Fluor™ 488 Imaging Kit » (Thermo Fisher Scientific, Waltham, USA) was used according to the manufacturer's recommendations to visualize EdU-incorporating cells. Cells were examined under a Zeiss LSM710 confocal microscope. Instrument settings were adapted to obtain the best possible resolution of the images by calculation of the theoretical resolution considering optical laws. Image J software (Image J 1.42 NIH, USA) was used for image treatment. Three fields were analyzed near the wound and three other ones at a distance from the wound.

## 2.8-MUC5AC immunolabelling, confocal microscopy observations and image analysis

After NP treatment, cells were fixed using 4 % paraformaldehyde in HBSS (Sigma-Aldrich) for 25 min at RT, and then permeabilized in 1 % BSA - 0.2 % Tween 20 (Sigma-Aldrich) in PBS for 10 min at RT. Fixed cells were incubated over night at 4 °C with human MUC5AC monoclonal primary antibodies (Thermo Fisher Scientific [MA5-12178]) diluted to 1:250 in the saturation-permeabilization solution. After 3 rinses with PBS, fixed cells were incubated with secondary antibodies stained with Alexa 488 diluted to 1:1000 in the saturation-permeabilization solution (Thermo Fisher Scientific [A-10667]) for 2 h at RT. Mounting was done in FluoroShield® (Sigma-Aldrich containing DAPI (4',6-diamidino-2-phenylindole, dihydrochloride, final concentration 0.25 μg/mL in PBS, Life Technologies). Cells were observed using a Zeiss LSM710 confocal microscope equipped with objective 63 X and a 1.5 zoom and images were treated with Image J software (Image J 1.42 NIH, USA). After a homogeneous threshold of confocal images, automatic image-cutting of areas corresponding to MUC5AC labelling was done. Their surface area was calculated after lens calibration. The whole surface area corresponding to the labelling was normalized to the number of nuclei identified by DAPI staining. Z-stack merges were obtained with the merge of 5 to 10 z-stacks with a stack-spread of 0.45 μm.

## 2.9-MUC5AC mRNA expression by quantitative reverse transcription polymerase chain reaction

After 24 h of treatment, cells were rinsed with HBSS and total RNA was extracted using NucleoSpin® RNA (Macherey-Nagel, Hoerdt, France) following the manufacturer's recommendations. RNA was quantified by Nanodrop™ (Nanodrop™ 2000, ThermoFisher Scientific). Reverse transcription (RT) of 1 μg of total RNA was performed using the High-Capacity cDNA Reverse Transcription Kit (ThermoFisher Scientific, France). Quantitative polymerase chain reaction (PCR) was performed with a Roche LC480 using LightCycler®480 SYBR Green 1 Master (Roche Diagnostics, Mannheim, Germany). cDNA was diluted to 1/10 in deionized water and amplified using specific primers and the following conditions: 5 min at 95 °C; 45 cycles of 10 s at 95 °C, 20 s at 58 °C and 20 s at 72 °C. The cycle threshold (CT) values were determined and normalized by the ∆∆CT method with RPL19 as housekeeping gene, using LightCycler®480 sofltware1.5. The primers for *MUC5AC* are: forward : GGCAACACCCTCCTCTAGCA, reverse : ACCGTGGAAGGCTCTGTGAT.

## 2.10-Single particle ICP-MS

An HR-ICP-MS Element II (Thermo Fisher Scientific, Germany) was used for the analyses. To resolve the main interference (*ca.* $^{40}Ar^{16}O$ at m/z = 56), $^{56}Fe$ was monitored in medium resolution mode (MR, R > 4000) as the separation power of the Element II in MR is high enough. In order to have the best stability, sensitivity and the lowest oxide rates (UO/U typically < 8 %), the HR-ICP-MS was optimized daily. spICP-MS measurements acquired 45,000 data with a 1 ms dwell-time. For the calculation of efficiency, $^{197}Au$ was monitored using the same resolution, dwell time and flow-rate as for $^{56}Fe$. This latter was manually calculated in triplicate (*ca*. 0.2 mL/min). Before the measurement of unknown samples, a suspension diluted $10^5$ times (NP number 2.60 x$10^5$ particles/mL) was freshly prepared to calculate transport efficiency using the size method, based on the TEM-measured median size of 60 nm particles. Citrate capped AuNPs (60 nm nominal diameter) were from British Biocell International (Cardiff, UK). The mean diameter provided by the supplier was 59.9 nm with a coefficient of variation of 8 % and the concentration was 2.60 x $10^{10}$ particles/mL. At millisecond dwell times the probability of coincidence decreases by decreasing the number concentration. Thus, several NP concentrations were studied and the preferred concentration were the highest without particle coincidence. All NPs and samples were diluted in ultrapure water. The limit of quantification of dissolved iron was 25 ng/L and the minimum detectable mass of 1.75 x $10^{-8}$ ng corresponded to an equivalent diameter for $Fe_2O_3$ spheres of 18 nm (*i.e.* the size limit of detection). Dissolved iron concentration was also monitored in blanks and EMEM culture medium for background correction.

**2.11-Statistical analysis**

Every experiment was repeated at least 3 times with triplicates for each condition. Data are represented as means ± standard error of the mean and were analyzed on commercially available software GraphPad Prism (Version 73.0, Systat software IncGraphPad Software, La Jolla, California, USA) using analysis of variance, one- or two-way ANOVA, followed by different post-hoc tests according to the type of experiment. The type of test is specified in the caption of each figure.

# 3 - Results and discussion

BWPs used in this study were derived from different braking systems and driving/testing conditions and provided a sample representative of the exposure to a wide diversity of brakes currently used in France. They were fully characterized in our previous study, as was the nanometric fraction (BWNPs) that we investigated here (figure S2) (Puisney et al., 2018). BWNPs were used in comparison to $\gamma$-$Fe_2O_3$ NPs (figure S3) as they are characterized by a high iron content and $\gamma$-$Fe_2O_3$ NPs were shown to be present in brake wear (Kukutschová et al., 2011). Moreover, this constitutes another example of a particulate xenobiotic with a well-defined particle size. The Calu-3 cell line was chosen as cellular model because it has the ability to differentiate *in vitro* by producing mucins and to form a tight barrier as shown in figure S4 (George et al., 2015b). Calu-3 cells can grow either in submerged conditions or at the air-liquid interface to better mimic the *in vivo* situation (Sanchez-Guzman et al., 2021). Here, the only possibility was to perform submerged cultures as BWNPs were recovered in a diluted aqueous suspension, meaning it was not possible to study pseudo-air-liquid interface exposure (Boublil et al., 2013). A wide range of concentrations was used as a compromise both to cover realistic concentrations and have an excess concentration to test the sensitivity of the model and of the evaluated endpoints.

*3.1-BWNPs exposure did not alter barrier integrity or promote a pro-inflammatory response*

Figure 1 shows that BWNPs did not increase the permeability of LY, a marker of the integrity of the epithelial barrier, whatever the concentration used. $\gamma$-$Fe_2O_3$ NPs showed a significant effect compared to the control from 5 µg/cm² and induced LY permeability higher than 2% from 25 µg/cm². However, this increase did not exceed 5% and confocal observations of the immunolabeling of the tight junctions-associated protein (ZO-1) did not show major alteration in the expression of this marker (Figure S5). The absence of alteration of the barrier integrity was also supported by the assessment of cytotoxicity with LDH release, which showed no increase in the treated cultures compared to the control cultures (Figure S6). This concentration range was kept for the rest of the study. BWPs have been shown to induce a pro-inflammatory response (Barosova et al., 2018) (Gasser et al., 2009), which is why Elisa assays were performed for 3 cytokines (IL-6, IL-8, TNFα) on apical and basal media (Figure S7). The absence of endotoxin contamination on NPs was checked (Table S1) as it could be responsible for the pro-inflammatory response. Both BWNPs and γ-$Fe_2O_3$ NPs had only a slight significant effect on TNFα release, but this effect was reversed. BWNPs had a limited induction effect (10 to 15% increase) at the two highest concentrations, whereas γ-$Fe_2O_3$ NPs reduced TNFα release by 11 to 13% according to the concentration from 25 µg/cm² (Figure S7). These results are in line with those obtained with BWNPs on 2D Calu-3 cells showing a very slight increase of IL-6 release from 25 µg/cm² (Puisney et al., 2018).

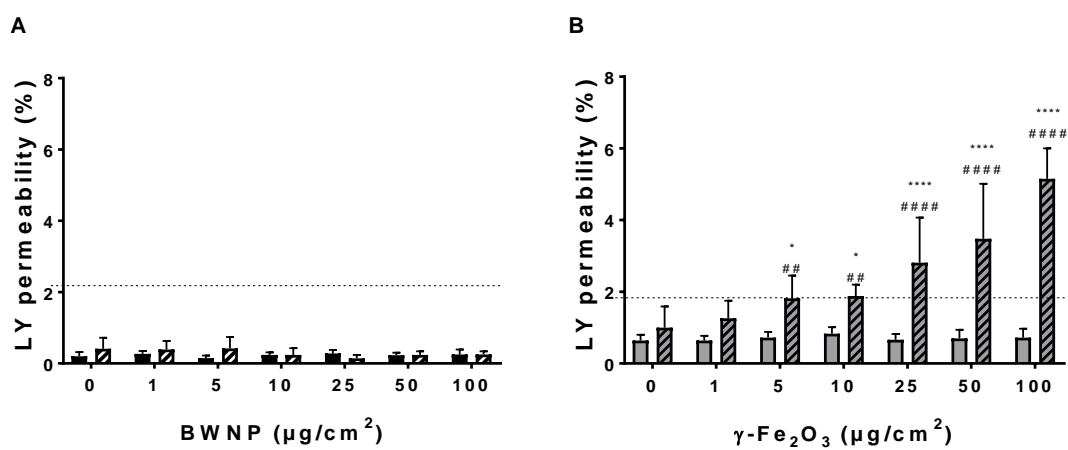

*Figure 1: Effect on Calu-3 epithelial barrier integrity of a 24 h exposure to BWNPs and γ-$Fe_2O_3$ NPs. Calu-3 epithelial barrier was exposed or not to BWNPs (A) and γ-$Fe_2O_3$ NPs (B) from 1 to 100 µg/cm². Lucifer yellow (LY) permeability was measured before (first bar for each concentration) and after 24 h exposure (second bar for each concentration). The dotted line represents the limit below which the integrity of the barrier is not preserved ((Sanchez-Guzman et al., 2021). Data are mean ±SD, two-way Anova (Sidak test); n = 6; N = 3; \*: significantly different from the control condition after 24 h exposure, #: significantly different between exposure for 0 and 24 h. \* : p>0.05 ; \*\* : p<0.01 ; \*\*\* : p<0.001 ; \*\*\*\* : p<0.0001.*

### 3.2-BWNPs translocate across the epithelial barrier in minute quantities

The signal obtained in spICP-MS consists of a series of intensities measured as a function of time and corresponding to the number of ions arriving at the detector after atomization and ionization in the plasma. Therefore, the background signal corresponds to dissolved iron concentration, whereas pulses represent the cloud of ions produced when a nanoparticle is atomized. Iron ($^{56}$Fe) was measured here as it is the main component of BWNPs, and accounts for 29% of all atoms detected by x-ray fluorescence (Puisney et al., 2018). spICP-MS measurements were performed on iron oxide (maghemite, γ-$Fe_2O_3$) NPs that were synthesized by the Massart route (Massart et al., 1995). Used as an analogue of BWNPs, these model NPs enabled us to quantify the amount of Fe in the apical and basal regions of air-liquid interface cell cultures, and thus to calculate the translocation rates through the epithelial support membrane. Representative examples of spICP-MS intensities obtained for BWNPs are illustrated in Figure S8. Figure 2a displays the integrated intensities as a function of concentration for the two types of NPs, here dispersed in the cell culture medium. It shows that the intensities increase linearly versus concentration (as indicated by the straight lines), the data for maghemite being around 100 times larger than that of BWNPs.

This difference can be explained by the fact that x-ray fluorescence does not take into account atoms of standard atomic mass below 22, i.e. including carbon, oxygen and nitrogen, which are abundant in BWPs (Grigoratos and Martini, 2015). Figures 2b and 2c show the changes in iron molar concentration in the apical and basal compartments for $\gamma$-Fe$_2$O$_3$ NPs and BWNPs, respectively. Each data point is averaged over 7 to 9 different dispersions exposed to cells, the error bars representing the standard deviation.

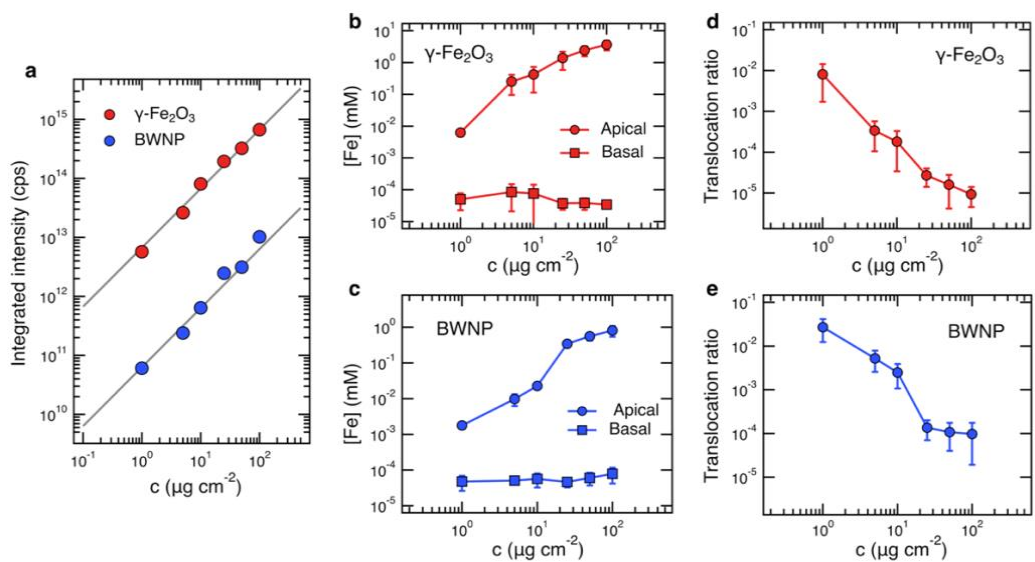

**Figure 2: Study by spICP-MS of translocation of NPs across the epithelial barrier.** *a) Integrated intensity indicated by single-particle ICP-MS as a function of the concentration of $\gamma$-Fe$_2$O$_3$ nanoparticles and BWNPs. The solid lines represent best fit calculations using a linear dependence. The difference between the two data sets is explained in the main text. **b and c)** Molar concentration of particulate iron indicated by single-particle ICP-MS as a function of the nanoparticle concentration dispersed in the apical region for $\gamma$-Fe$_2$O$_3$ nanoparticles and BWNPs, respectively. **d and e)** Translocation ratios for the particles described in b) and c).*

The main features here are the steady and almost linear iron concentration increase in the apical region and its relative constancy in the basal part, which lies around of $10^{-4}$ mM. Note that the values for BWNPs are typically 5-10 times lower than for maghemite model particles. The ratio between the basal and apical concentrations, also called the translocation ratio, is calculated and shown in Figures 2d and 2e, respectively. In both cases this ratio shows a strong decrease, indicating that whatever the exposure concentration, the number of particles crossing the epithelial barrier remains limited. This last result can be understood by invoking a feature shared by numerous NPs in cell culture media, namely their aggregation (Falahati et al., 2019). Studies have also shown, especially on bare iron oxide (Safi et al., 2011) (Giamblanco et al., 2017) and on BWNPs (Puisney et al., 2018) that the aggregate size increases drastically with the nanoparticle concentration. In certain cases, the aggregates can be a few microns across, and can not be translocated through the pores of the Transwell membranes. Taken together, the spICP-MS results in Figure 2 show low translocation ratios, around 0.1-1 % for c = 1-10 µg/cm$^2$ and $10^{-3}$-$10^{-2}$ % in the range 10-100 µg/cm$^2$, with both the model iron oxide particles and the BWPs. These results are in agreement with our previous study using the Calu-3 barrier where, using fluorescently labelled NPs, we showed a translocation whatever the NP composition (SiO$_2$ or TiO$_2$), but translocation was increased for the smallest and negatively charged NPs (George et al., 2015a). Although extrapolation to what might happen *in vivo* would probably be too speculative, it can nevertheless, be noted that for other types of particles of similar size, these low translocation rates have been reported in animal instillation studies (Möller et al., 2008) (Kreyling et al., 2014).

### 3.3-BWNPs induced MUC5AC expression by the EGFR-dependent pathway

The mucus layer covering bronchial epithelial cells has a protective role by allowing the mucociliary clearance by ciliated cells. Its production must be finely controlled to avoid overproduction, which would alter mucociliary activity and increase proneness to infection. In order to investigate whether BWNP exposure modulates mucus production, MUC5AC, one of the main mucins present in the bronchial mucus, was studied both at its transcription and protein level. BWNP exposure induced a 5-fold increase in of MUC5AC expression compared to control culture at 100 µg/cm² (Figure 3A). Immunolabeling of MUC5AC was performed to evaluate protein production and confocal microscopy showed an increase in the number of vesicles corresponding to mucus granules as BWNP concentration was increasing with both more and more cells involved, and more vesicles in each individual cell (Figure S9). Quantification of MUC5AC-immunolabelled surfaces showed a significant increase in the fluorescence level from 5 µg/cm² (Figure 5C). At 100 µg/cm², the fold induction was 5 and therefore consistent with that observed in the MUC5AC transcript study. The same observations were made for exposure to γ-$Fe_2O_3$ NPs, except that the induction factor was 8 (Figure 3B-D, figure S9). Altogether, these data highlight the ability of BWNPs as well as γ-$Fe_2O_3$ NPs to concentration-dependently induce mucus production and are consistent with findings obtained using other metal NPs, Cu NPs, in *in vitro* and *in vivo* experiments (Ko et al., 2016) (Park et al., 2016).

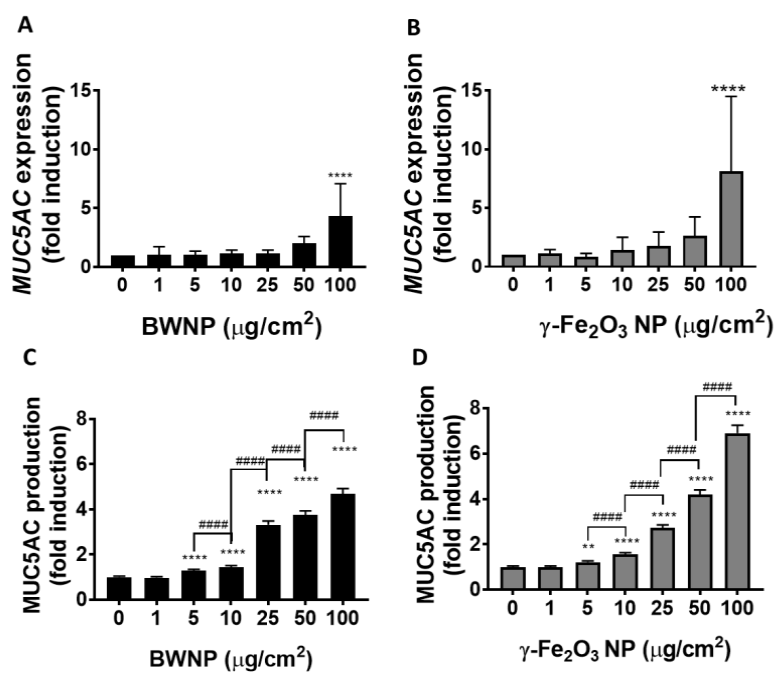

*Figure 3: MUC5AC expression at the mRNA and protein level in Calu-3 cells exposed for 24 h to BWNPs and γ-$Fe_2O_3$ NPs at different concentrations (µg/cm²). (A-B) At the end of the incubation, mRNA was extracted and RT-qPCR was performed to evaluate MUC5AC expression. Results are mean ± SEM and expressed as fold induction compared to the control. (\* : different to the control and the other concentrations; \*\*\*\* : <0.0001 ; one-way ANOVA (Tukey post-test); n = 9; N = 3 ;). (C-D) At the end of the incubation, cells were fixed and immunolabeled to identify MUC5AC. MUC5AC labelling observed by confocal microscopy was quantified by image analysis using ImageJ. One-way ANOVA Tukey post test; N*

*= 5 fields of n = 22 +/- 4 cells; Mean ± SD; \* : different to the control, # : different between the two concentrations shown with brackets; \*\* : p<0.01; \*\*\*\*/#### : p<0.0001)*

The EGFR pathway is a signalling pathway involved in the regulation of mucus production following exposure to xenobiotics such as cigarette smoke (Takeyama et al., 2001). The EGFR pathway can be activated by a mechanism that is ligand- or non-ligand-dependent (Figure S9) (Burgel, 2004). Non-ligand EGFR activation can be triggered by reactive oxygen species (Goldkorn et al., 2014) and we previously showed that BWNP exposure increased levels of reactive oxygen species in 2D Calu-3 cells (Puisney et al., 2018). The role of oxidative stress in the toxicity mechanisms of BWNPs is explained by their content of redox-active metals and it has been found that brake wear debris has a significant oxidative potential (Rajhelová et al., 2019). To determine whether the EGFR pathway contributes to MUC5AC induction by BWNPs and the underlying mechanism, we used a neutralizing antibody (anti-EGFR) that prevents ligand binding as well as an EGFR antagonist (AG1478) (Figure S10). As shown in Figure 4, in the presence of AG1478, MUC5AC expression both at the mRNA and protein level was completely abolished, showing the major role of the EGFR pathway in the induction of MUC5AC by the two types of metal NPs. The specificity of this effect was checked using AG9, the inactive form of AG1478, which was inefficient in inhibiting NP-induced MUC5AC expression/production.

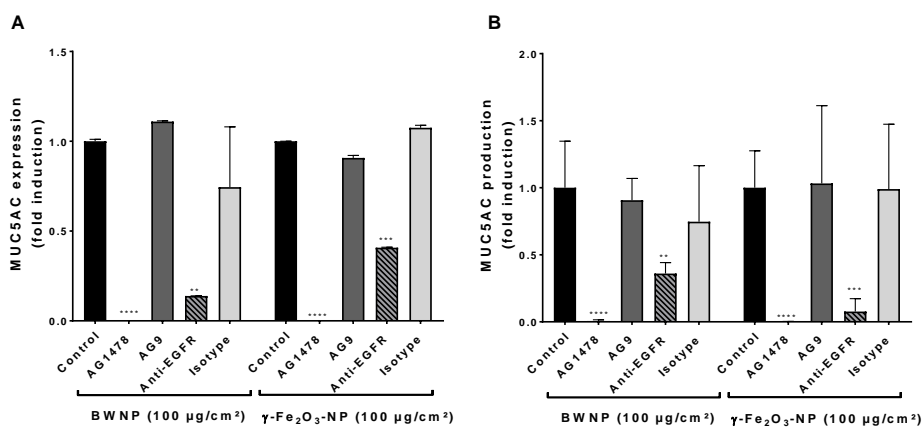

*Figure 4: MUC5AC expression and production in Calu-3 cells exposed for 24 h to BWNPs and γ-$Fe_2O_3$ NPs at 100 µg/cm² in the presence of different inhibitors. Cells were incubated for 30 min with the inhibitor/antibody before being exposed for 24 h to NPs at 100 µg/cm² (A) MUC5AC mRNA expression. Results are mean ± SEM and expressed as fold induction compared to the control. \* : different to the control; \*\*\*\* : <0.0001 ; two-way ANOVA (Dunnett post-test) ; n = 6 ; N = 3 . (B) quantification of MUC5AC immunolabeling. Two-way ANOVA (Dunnett post test) ; N = 5 fields of n = 22 +/- 4 cells ; Mean ± SD; \* : different to the control, # : different between concentrations ; \*\* : p<0.01; \*\*\*\*/#### : p<0.0001)*

A anti-EGFR induced a significant reduction in MUC5AC expression at both the mRNA and protein levels. The specific action of the neutralizing antibody was checked using a non-specific antibody that did not share this property of reducing NP-induced MUC5AC expression/production. These data suggest that EGFR pathway activation is partly dependent on ligand binding. There are different EGFR ligands and two of them (amphiregulin and TGFα) were previously shown to be involved in mucus production by bronchial cells exposed to PM2.5 (Val et al., 2012) (Wang et al., 2019). Amphiregulin release by Calu-3 cells was not modified by 24 h exposure to either BWNPs or γ-$Fe_2O_3$ NPs (Figure S11A) whereas TGFα induced a significant but limited reduction in its release (22% at 1 µg/cm²), but in a non-concentration-dependent manner (Figure S11B). The fact that there was no increase in ligand release while the EGFR ligand-dependent pathway was shown to participate in MUC5AC expression could be explained by a fast autocrine/paracrine action of released ligands or the involvement of other EGFR ligands. The mucus hyperproduction induced by this acute exposure could be seen as protective, by reducing NP uptake by epithelial cells, but in the case of chronic exposure it could induce a shift in hypersecretory phenotype, which is a hallmark of chronic obstructive diseases such as asthma and chronic obstructive respiratory disease (Ma et al., 2018).

### 3.4-Preexposure of the epithelial barrier to BWNPs did not alter the rate of repair, but stimulated the proliferation rate at the wound healing edge

Due to its position as an interface, the respiratory epithelium is frequently subject to injuries requiring an efficient repair. This consists of several steps, including a migration of cells adjacent to the wound to fill the mostly quiescent. By contrast, for cells at the edge of the wound, EdU incorporation became significant from 6 h, with 10% of nuclei labelled, and increased until 16 h, with 38% of nuclei labeled (Figure S13). However, cells further away from the wound, exhibited 24% of labelled nuclei as early as 1 h after wounding, and this percentage rose to 60% at 16 h. This high rate of EdU incorporation in this part of the epithelium, was not modified by a preexposure to BWNPs or $\gamma$-$Fe_2O_3$ NPs (Figure 5B, 5D, respectively). By contrast, at the edge of the wound the incorporation of EdU was increased in BWNP-treated cells as early as 1 h at 10 and 100 µg/cm² (Figure 5A, figure S14). Whatever the time point there was a significant difference from control culture at 100 µg/cm² (except at 10 h). The same conclusions can be drawn for $\gamma$-$Fe_2O_3$ NPs but to a lesser extent (Figure 5C, figure S14). The increase of EdU incorporation reflects an increase in the number of cells engaged in a proliferative process. The incorporation of EdU, beyond the proliferation mechanism, could also indicate DNA repair induced by NP exposure. Nevertheless, a significant difference in the epithelia exposed to NPs was only observed in the wounding zone. If DNA damage occurred in the cells during exposure, it should also have occurred in the cells far from the injury. We did not find in the literature similar experiments performed with NPs. Nevertheless, very small iron oxide nanoparticles used as contrast agents transiently inhibited the proliferation of human blood outgrowth endothelial cells when used at high concentrations (Soenen et al., 2010). Altogether, these data suggest that in cultures pre-exposed to NPs, the onset of wound filling involves both migrating and proliferating cells and that cell cycle activation in the vicinity of the lesion is larger.

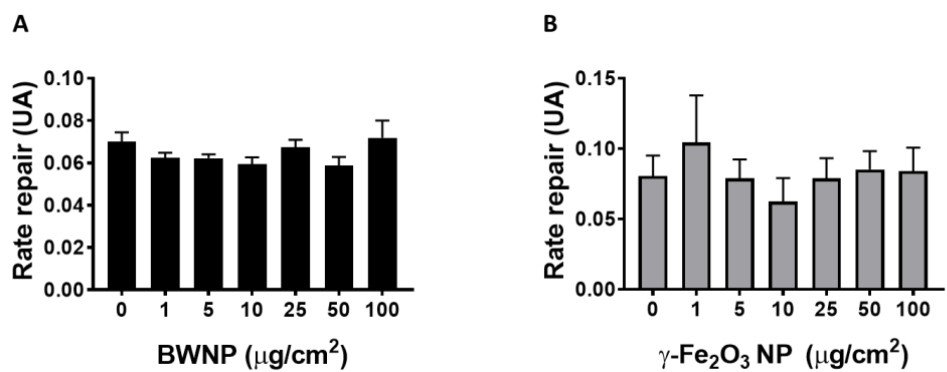

*Figure 5: Repair rate of wounded Calu-3 cell cultures previously exposed to BWNPs and $\gamma$-$Fe_2O_3$ NPs at different concentrations (µg/cm²). After 24 h exposure to particles, a straight wound was made in the cell culture and cells were labelled with CellTracker™ Green CMFDA to view them with a full-field microscope. (A-B) quantification of the repair process according to the concentration of A) BWNPs and B) $\gamma$-$Fe_2O_3$ NPs. Values are mean ± SD; one-way ANOVA, Kruskall-Wallis, N = 3, n = 3, UA = % of repair/min.*

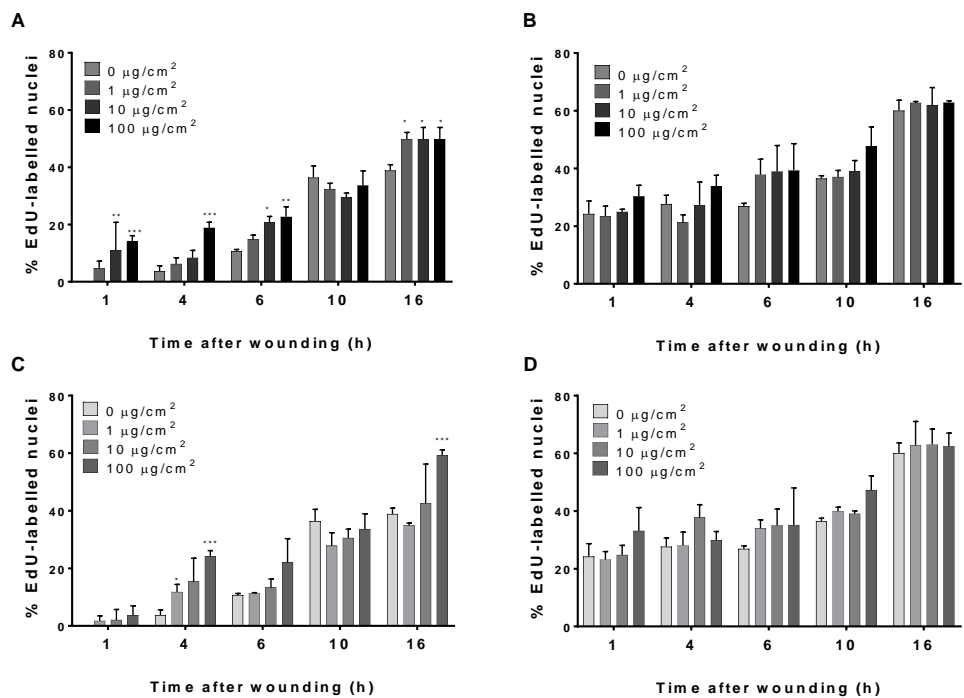

*Figure 6: EdU incorporation during the repair of Calu-3 cell epithelia after wounding preceded by 24 h exposure to different NP concentrations. After 24 h exposure to BWNPs (A-B) and ɣ-Fe$_2$O$_3$ NPs (C-D), a straight wound was made in the cell culture and cells were incubated with EdU for 1 to 16 h. Level of EdU incorporation as a function of time was determined after quantitative analysis of images acquired by confocal microscopy either at the wound healing edge (A-C) or at the periphery of the culture (B-D). Two-way ANOVA Tukey post test; N = 3 fields of n = 19 +/- 4 cells; Mean ± SD; Objective: 40X - oil immersion; \*: different from control (0) of the corresponding time point; \* : p<0.05 ; \*\*\* : p<0.001)*

## 3.5 - Conclusion

In this study we provide evidence that the finest fraction of BWPs has the ability to cross the epithelial barrier, suggesting the possibility of systemic effects linked to NP accumulation in secondary organs. Although the translocation rate may appear low, it should be noted that the experiment was carried out following a single exposure, whereas in real life exposures is repeated, which in the long term can lead to significant concentrations in certain organs in the case of inefficient clearance. This acute exposure was also responsible for mucus overproduction, which can be considered as beneficial if transient, but can contribute to bronchial remodelling in long-term exposure. The evidence of an impact on cell proliferation in injured epithelium could be responsible for altered differentiation or even promote neoplastic processes. Thanks to this relevant culture model, all the data obtained indicate damage to the epithelial barrier after an acute exposure. This raises questions about the possible consequences of chronic exposure, which represents the reality of human exposure.

## Conflict of interest disclosure

The authors declare no competing financial interest.

## Credit author statement

Chloé Puisney-Dakhli: methodology, investigation, formal analysis, visualization, Evdokia Oikonomou: investigation, Mickael Tharaud: investigation, resources, Yann Sivry: conceptualization, resources, validation, Jean-François Berret: conceptualization, supervision, writing, funding acquisition, Armelle Baeza-Squiban: conceptualization, supervision, project administration, writing, funding acquisition.

**Acknowledgments**

The Ph.D. of Chloé Puisney-Dakhli at the University Paris-Diderot (now Université Paris Cité) was funded by the Région Ile-de-France and the DIM (Domaine d'Intérêt Majeur) NanoK. We acknowledge X. Baudin from the confocal microscopy platform, and the ImagoSeine BioImaging Core Facility (ANR-10-INSB-04, Investments for the future), Institute Jacques Monod, Paris, France. The authors wish to thank Olivier Lambert of Univ. Bordeaux, CBMN UMR 5248, Bordeaux INP for TEM images. Agence Nationale de la Recherche (ANR) and Commissariat à l'Investissement d'Avenir (CGI) are acknowledged for financial support of this work through Labex Science and Engineering for Advanced Materials and devices (SEAM) ANR 11 LABX 086, ANR 11 IDEX 05 02. This research was also supported in part by the ANR under the contract ANR-12-CHEX-0011 (PULMONANO), ANR-15-CE18-0024-01 (ICONS), ANR-17-CE09-0017 (AlveolusMimics) and by Solvay. The authors would also like to acknowledge the support by the IPGP multidisciplinary program PARI and by Region Île-de- France SESAME Grant no. 12015908.